\magnification 1200
\baselineskip=17pt

\centerline{\bf THEORY FOR DECOUPLING IN HIGH-T$_{\rm c}$ SUPERCONDUCTORS FROM}
\bigskip
\centerline{\bf AN ANALYSIS OF THE LAYERED XY MODEL WITH FRUSTRATION}
\vskip 50pt
\centerline{J. P. Rodriguez}
\medskip
\centerline{\it Dept. of Physics and Astronomy,
California State University, Los Angeles, CA 90032.}
\vskip 30pt
\centerline  {\bf  Abstract}
\vskip 8pt\noindent
The nature of decoupling in the mixed phase of extremely type-II layered
superconductors is studied theoretically through a duality transformation
of the layered $XY$ model with frustration.  
In the limit of weak coupling, we generally find that 
the 
 Josephson effect    is absent if and only 
if the  phase correlations within isolated
layers are short range.  
In the case specific to
uniform frustration, we notably identify a decoupled pancake vortex
liquid phase that is bounded by first-order and second-order decoupling
lines in the magnetic field vs. temperature plane.  These transitions
potentially account for the flux-lattice melting and for the flux-lattice
depinning that is observed in clean high-temperature superconductors.

\bigskip
\noindent
PACS Indices:   74.60.-w, 74.25.Dw, 74.25.Ha, 74.60.Ge


\vfill\eject

{\it Introduction.} 
High-$T_c$ superconductors 
are perhaps the best known example of an extremely type-II
layered superconductor.[1]
The vortex lattice that exists in
the mixed phase of clean oxide superconductors melts through a first-order
transition
for magnetic fields applied perpendicular to the layers.[2]
The first-order line,
$H_{\perp} = H_m (T)$, begins at the zero-field critical point, $T_c$,
but it ends strangely  in the middle of the phase diagram.
The depinning       line
$T = T_{dp}(H_{\perp})$, which marks the point at which the flux lattice
depins itself through   thermal excitations,
appears to be independent of   this melting line.[3,4]
The first-order transition is commonly interpreted as either a
vortex lattice melting transition,[1]
 or as a layer decoupling transition.[5,6]
The respective theoretical approaches have been primarily based on elastic
medium descriptions of the vortex matter that  notably omit
topological defect excitations.[7,8]
These theories   also fail  to   account for the multicritical point
mentioned above in the absence of disorder.[9]

In this paper, we shall model the thermodynamics of the interior of the
mixed phase in extremely type-II anisotropic superconductors by the
layered $XY$ model with uniform frustration.[10-12]  
The nature of the Josephson effect   between layers is uncovered 
through a {\it partial} duality analysis       of the $XY$ model.[13-17]
A subsequent weak-coupling  analysis at high perpendicular fields 
yields a low-temperature phase
made up of coupled 2D vortex lattices 
that is separated from a
decoupled liquid of planar vortices at high temperatures by a second
order melting line.  It is further  argued that the decoupled liquid 
phase experiences
a first-order phase transition into a coupled solid phase as 
the perpendicular field is lowered
into the strong-coupling regime.[16,17]  The phase diagram that results
(see Fig. 1)
is compared   with recent experimental reports of multicritical phenomena
in the mixed phase of high-temperature superconductors.[3,4]

{\it    Layered $XY$ Model.}  Before considering  the 
uniformly frustrated case that describes 
the mixed phase of extremely type-II superconductors, let us  
first perform  a duality analysis  of the layered $XY$ model without
frustration.[18]
The corresponding energy functional reads
$$\eqalignno{ E_{XY}^{(3)} = - J_{\parallel}&
\sum_{l=1}^{N}\sum_{\vec r}\sum_{\mu=x,y}
{\rm cos}[\Delta_{\mu}\phi(\vec r,l)-A_{\mu}(\vec r,l)]\cr
& - J_{\perp}\sum_{l=1}^{N-1}\sum_{\vec r}
{\rm cos}[\phi(\vec r, l+1)-\phi(\vec r, l)-A_z(\vec r,l)],
& (1) \cr}$$
where  $\phi(\vec r, l)$ is the superconducting
 phase at a point
$\vec r$ in layer $l$.
Above , $\Delta_{\mu}$ denotes the
nearest-neighbor difference operator along the $\hat\mu$ direction, 
 while $A_{\mu}$ represents a purely longitudinal vector potential.
The Josephson penetration length 
$\gamma^{\prime} a$
provides a natural scale for  the model (1)  in the limit of
weak inter-layer coupling,  in which case
the model anisotropy parameter 
$\gamma^{\prime} = (J_{\parallel}/J_{\perp})^{1/2}$ is  much larger
than unity.  Here, $a$ denotes the (square) lattice constant for
each layer.   Last, 
any generalized phase auto-correlation function
set by an     integer source field, $p(r)$, is related to the
corresponding partition function
$$Z_{XY}^{(3)}[p]=
\int {\cal D} \phi\,  e^{-E_{XY}^{(3)}/k_B T} e^{ i\sum p  \phi}\eqno (2)$$
by the quotient
$$\Bigl\langle {\rm exp} \Bigl[i\sum_r p(r) \phi(r)\Bigr]\Bigr\rangle = 
Z_{XY}^{(3)}[p]/Z_{XY}^{(3)}[0].\eqno (3)$$
We shall now employ the well-known dual representation[13]
of the $XY$ model (1) based on the Fourier series expansion    
$e^{\beta{\rm cos} \theta} =
 \sum_{n = -\infty}^{\infty} I_{|n|} (\beta) e^{in\theta}$
 of  the Gibbs
distribution in terms of modified Bessel functions, $I_n (x)$.  
This identity allows the phase variables to be integrated out of (2).  
Resummation  over the parallel link fields
$n_x$ and $n_y$ then yields the form
$$Z_{XY}^{(3)}[p]  = I_0^{{\cal N}^{\prime}}(\beta_{\perp}) \cdot
Z_{\rm CG}[p]\cdot \Pi_{l = 1}^N Z_{XY}^{(2)}[0] \eqno (4)$$
for the partition function of the layered $XY$ model
in terms of a product of a  layered  Coulomb gas ensemble (CGE)
$$Z_{\rm CG}[p] = \sum_{\{n_{z}(r)\}} y_0^{N[n_z]}\cdot
\Pi_{l = 1}^N C_l   [q_l]\cdot
e^{-i\sum_r n_z A_z}   \eqno (5)$$
with $N$ isolated   $XY$ model layers ($J_{\perp} =  0$).[17] 
Above, $n_{z}(r)$ is  the remaining integer 
link field between adjacent layers
that  is effectively restricted to
take values $n_z(r) = 0, \pm 1$ 
in the weak-coupling limit 
$J_{\perp}\ll k_B T$. 
Indeed,     each configuration in the CGE (5) is weighted
by the bare fugacity
$y_0 =  J_{\perp}/2 k_B T$
raised to the power
$N[n_z] \cong  \sum_{\vec r,l} |n_z(\vec r,l)|$  equal to the total
number of $n_z$-charges.  The latter so-called fluxons physically
represent vortex rings in between adjacent layers.[14]
Each configuration is further weighted by a product of
generalized auto-correlation functions
$C_l [q_l] = \Bigl\langle {\rm exp} \Bigl[ i \sum_{\vec r}
q_l(\vec r) \phi(\vec r,l)\Bigr]\Bigr\rangle_{J_{\perp} = 0}$
over each layer $l$ in isolation that are
 evaluated at sources
$q_l(\vec r) =  p(\vec r, l) + n_z(\vec r, l-1) - n_z(\vec r, l)$.
Last, we   have the parameters 
$\beta_{\parallel, \perp} = J_{\parallel, \perp}/k_B T$, while  
${\cal N}^{\prime}$ denotes
the total number of rungs between layers.

To proceed further, it is instructive to consider
a single  neutral  pair of unit $n_z$ charges
that lie  in between
layers $l^{\prime}$ and $l^{\prime}+1$ in the absence of an external
source, $p = 0$, with the
negative and positive charges located at planar sites
$\vec r_1$ and $\vec r_2$, respectively.
The gauge-invariant product over intra-layer autocorrelation functions 
in the layered CGE
 (5) then reduces
to the product 
$|C_{l^{\prime}} (\vec r_1-\vec r_2)| \cdot
|C_{l^{\prime}+1}(\vec r_1-\vec r_2)|$
of the corresponding  
phase autocorrelation functions,
$$C_{l} (\vec r_1, \vec r_2) = \Bigl\langle {\rm exp}
\Bigl[i\phi (\vec r_1,l) - i\phi (\vec r_2,l)
\Bigr]\Bigr\rangle_{J_{\perp} = 0}\ ,
\eqno (6)$$
within isolated layers.
This function 
varies algebraicly with the separation   as
$$|C_l (\vec r)| = g_0 
(r_0/|\vec r\,|)^{\eta_{2D}} 
\quad {\rm for} \quad |\vec r|\ll \xi_{\rm vx}, \eqno (7)$$
while it decays exponentially with the separation  as
$$|C_l (\vec r)| = g_0 \, {\rm exp}
(-|\vec r\,|/\xi_{\rm vx})
\quad {\rm for}\quad |\vec r|\gg \xi_{\rm vx}.\eqno (8)$$
Here, 
$\eta_{2D} = \eta_{\rm sw} + \eta_{\rm vx}$
is the 2D correlation exponent inside layer $l$,  where 
$\eta_{\rm sw} = (2\pi\beta_{\parallel})^{-1}$
and  $\eta_{\rm vx}$ are the  respective
spin-wave and vortex contributions.
Also,
$\xi_{\rm vx}$ denotes the 2D
phase correlation length, while   the length
$r_0 = a / (2^{3/2} e^{\gamma})$ is  set by
Euler's constant, $\gamma$.
The effective layered CGE 
(5) therefore takes   the form
$$\eqalignno{ Z_{\rm CG} [0] =  \sum_{\{n_z\}}  
(g_0 y_0)^{N[n_z]}{\rm exp}
\Biggl\{  - {1\over 2}\sum_{l}
\sum_{\vec r_1, \vec r_2}^{\qquad} q_l(\vec r_1)
\Bigl[ \eta_{2D}
 & {\rm ln} (r_0/|\vec r_1 - \vec r_2|)
 - V^{[q_l]}_{\rm string} (\vec r_1,  \vec r_2)\Bigr]
q_l(\vec r_2) \cr
&-i\sum_{l}\sum_{\vec r}^{\qquad} n_z(\vec r, l) A_z(\vec r,l)\Biggr\}
 &(9)\cr}$$
in the limit of dilute fluxon ($n_z$) charges.
At separations  within a correlation length, 
$|\vec r_1 - \vec r_2| \ll \xi_{\rm vx}$,
the fluxons experience
a pure  Coulomb interaction ($V_{\rm string}^{[q_l]} = 0$).
At large separations $|\vec r_1 - \vec r_2|\gg\xi_{\rm vx}$,
on the other hand,
the fluxons experience
a pure ($\eta_{2D} = 0$) confining interaction
$V_{\rm string}^{[q_l]} (\vec r_1,\vec r_2) =
   |\vec r_1  -  \vec r_2|/\xi_{\rm vx}$
between those  points      $\vec r_1$ and $\vec r_2$
in layer $l$
that are connected by a  string [see Eq. (8) and ref. [19]].  
Below, we determine the thermodynamic 
nature of the superconducting and ``normal''    phases that correspond
to  Eqs. (7) and (8), respectively.

Consider first  the case where (quasi) long-range intra-layer
phase correlations are present: $\xi_{\rm vx} = \infty$ and
$V^{[q_l]}_{\rm string} = 0 $.  
Summing   independently over charge configurations of the CGE (9) 
that are restricted to take
values $n_z = 0, \pm 1$ at each site
plus
an appropriate Hubbard-Stratonovich transformation[12] yields
an equivalent
renormalized Lawrence-Doniach (LD) model[1]  set
by the continuum  energy functional 
$$\eqalignno{
E_{\rm LD} = &
\bar J_{\parallel}\int d^2 r \Biggl[
\sum_{l = 1}^{N}
{1\over 2}(\vec\nabla\theta_l)^2
-\Lambda_0^{-2}
\sum_{l = 1}^{N-1}{\rm cos}(\theta_{l+1}
-\theta_l - A_z)\Biggr].
&(10)\cr}$$
 Here,
$\bar J_{\parallel} = k_B T / 2\pi \eta_{2D}$
is the  macroscopic  phase rigidity of an isolated
layer,[20]  while	
$\Lambda_0 = a (\bar J_{\parallel}/ g_0  J_{\perp})^{1/2}$
is the renormalized Josephson scale.
The above continuum
description (10) is understood to have an  ultra-violet cut-off
 on the order of the lattice constant, $r_0   \sim a$.  
Eq. (10)  is known to sustain a macroscopic Josephson
effect  at temperatures below
$k_B T_{*} \cong 4\pi \bar J_{\parallel}$.[14,15] 
Also, the factorization (4) plus the  continuum limit  (10) indicate 
the expression
$$\langle {\rm cos}\,\phi_{l,l+1}\rangle \cong  
{1\over 2} \beta_{\perp}
 +  f_0   \Biggl({r_0\over{\Lambda_{\rm J}}}\Biggr)^{\eta}
 \eqno (11)$$
for the local Josephson coupling, 
where 
$\phi_{l,l+1}(\vec r) = 
 \phi (\vec r,l+1)- \phi (\vec r,l)-A_z(\vec r)$
is the gauge-invariant phase difference between consecutive layers.
An analysis of the double-layer case yields an effective anisotropy
scale
$\Lambda_{\rm J}
= \Lambda_0/2^{1/2}$ and
the limiting values
$\eta \rightarrow  \eta_{2D}$ and 
$f_0\rightarrow g_0$ for the exponent and for the prefactor
at low temperatures, $\eta_{2D}\ll 1$ (see ref. [17]).
Expression (11) indicates that
a crossover to strong coupling,
$\langle {\rm cos}\,\phi_{l,l+1}\rangle \sim 1$,
must therefore take place in the isotropic regime $a \sim \Lambda_0$
and/or at a temperature of order the Josephson 
scale $T_{\rm J} = J_{\perp}/k_B$.

Consider next the case in which intra-layer
correlations are short range:  $\xi_{\rm vx} < \infty$.  
Inter-layer fluxon ($n_z$) pairs are then
bound by a confining string (8).
Application of the CGE (5)
yields Koshelev's formula[11]   
$$\langle e^{i\phi_{l,l+1}}\rangle \cong  y_0
\int d^2 r
|C_l(r)| \cdot  |C_{l+1}  (r) |/a^2\eqno (12)$$
for the local Josephson coupling in this decoupled phase
(see ref. [17]).
Substitution of  the form (8) for the
short-range intra-layer autocorrelator 
in turn yields the explicit formula
$$\langle {\rm cos}\,\phi_{l, l+1}\rangle
\cong  {\pi g_0^2\over 4}
 \Biggl({\xi_{\rm vx}\over{a}}\Biggr)^2
\beta_{\perp}\eqno (13)$$
for the local Josephson coupling.  It can be shown that
the macroscopic Josephson effect   is absent in this decoupled
phase.[16,17]

We can now determine the phase diagram of the layered 
$XY$ model.[18]  It is well known that an isolated 
$XY$ layer looses phase coherence
 above a temperature $k_B T_c^{(2D)} \cong {\pi\over 2} J_{\parallel}$.[13]
The transition is driven by the unbinding of
vortex/anti-vortex pairs and it  is second-order.[21]
Consider first the weak-coupling limit
$\langle {\rm cos}\,\phi_{l,l+1}\rangle \rightarrow 0$.
By the previous  analysis, we then conclude that the layers
are Josephson coupled at low temperature $T < T_c^{(2D)}$ following
the renormalized LD model (10), while
they are decoupled at high temperature $T > T_c^{(2D)}$.
Eq. (13) indicates, also, that the selective high-temperature expansion
breaks down 
($\langle {\rm cos}\,\phi_{l,l+1}\rangle \sim 1$)
 in the decoupled phase at a temperature $T_{\times}$
set by the identification of length scales 
$\Lambda_0\sim\xi_{\rm vx}(T_{\times})$.
The second-order transition therefore
takes place
at a temperature $T_c$ that lies
inside of  the dimensional crossover window
$T_c^{(2D)} < T < T_{\times}$ for large yet finite  anisotropies
$\gamma^{\prime}\gg 1$.  These conclusions agree with what is
presently understood for the layered $XY$ model.[18]

{\it Uniformly Frustrated Case.}  Consider now the layered $XY$
model (1) in the presence of a uniform frustration,
$A_{\mu} = (0, b_{\perp} x, 0)$, which
describes the mixed phase of an extremely type-II 
superconductor in a field $B_{\perp} = (\Phi_0/2\pi a) b_{\perp}$
aligned perpendicular to the layers.[10-12]  The duality analysis
just performed can then be repeated wholesale,
yet  with the following modifications:[16,17]
({\it a}) Each isolated $XY$ layer now
undergoes a 2D melting transition
 that is mediated by the unbinding of dislocation pairs[21]
at a temperature[22] $k_B T_m^{(2D)} \cong J_{\parallel}/20$,
 while ({\it b}) the ultra-violet cut-off, $r_0$, of the renormalized LD
model (10) is now  of order the average spacing
between planar vortices, $a_{\rm vx} = (\Phi_0/B_{\perp})^{1/2}$.
(The CGE [Eqs. (7)-(9)] must be coarse-grained  
up to the new  ultra-violet scale, $a_{\rm vx}$.)
In the weak-coupling limit 
$\langle {\rm cos}\,\phi_{l,l+1}\rangle \rightarrow 0$
reached
at high fields $B_{\perp} \gg B_{\perp}^* = \Phi_0/\Lambda_0^2$,
we therefore have $N$ 2D vortex lattices that show a
macroscopic Josephson effect  at low temperatures $T < T_m^{(2D)}$,
while  a decoupled liquid
of planar vortices exists at high temperatures $T > T_m^{(2D)}$.  At
large yet finite anisotropies, $\gamma^{\prime} \gg 1$, it is useful
once again to determine  the temperature scale $T_{\times}$ at which
point the selective high-temperature expansion (13) for the local
Josephson coupling breaks down.  This again takes place 
roughly when $\Lambda_0$ and $\xi_{\rm vx}$ are comparable.
As in the previous case without frustration, we therefore expect a  
second-order melting transition at a temperature $T_m$ 
that lies inside the dimensional crossover window[18]
$T_m^{(2D)} < T < T_{\times}$.

Yet what happens when the local Josephson coupling 
$\langle {\rm cos}\,\phi_{l,l+1}\rangle$
 approaches unity as the field $B_{\perp}$ is lowered?
It is useful      to first define a decoupling contour
$$\langle {\rm cos}\, \phi_{l,l+1}\rangle =
\langle {\rm cos}\, \phi_{l,l+1}\rangle_D,\eqno (14)$$
in the $T$-$B_{\perp}$ plane.   Numerical simulations
indicate that
$\langle {\rm cos}\, \phi_{l,l+1}\rangle_D$ is
a constant less than but of order unity.[11]
The result (11) for the  local Josephson
coupling in the coupled phase yields ({\it i})
a contour line at temperatures
of order the Josephson energy,
$k_B T_{\rm J} = J_{\perp}$,  for high perpendicular
fields $B_{\perp} \gg B_{\perp}^*$,
and ({\it ii})
a contour  line at perpendicular fields of order $B_{\perp}^*$
for temperatures near     $T_m^{(2D)}$.
Since no phase transition is possible in the screened CGE (9) that
describes the coupled phase, this contour line must therefore represent
a crossover into a flux-line lattice regime that exists at
lower  temperatures and
fields. At high temperatures $T > T_{\times}$ 
inside the weak-coupling
regime of the decoupled phase, the string interaction (8)
binds together  fluxon-antifluxon pairs into  stable
dipoles of dimension $\xi_{\rm vx}$ that do not overlap.
It can be shown that such fluxon pairs begin to overlap 
and dissociate in the vicinity of the decoupling contour (14)
for temperatures  
that lie outside of the 2D critical regime 
($\xi_{\rm vx}\sim a_{\rm vx}$).[17]
We therefore expect a first-order transition along this line due to 
the absence of a nearby divergent length scale.

The above results are summarized by the 
schematic phase diagram
shown in Fig. 1.
The  phenomenology
$J_{\perp} \propto  (T_c-T)/T_c$
for the Josephson  energy
in the vicinity of the zero-field transition at $T_c$ yields
the same  linear temperature dependence
for the first-order decoupling field,[2,5,6]   
$H_{D} (T)$  [see Eqs. (13) and (14)].    
Eq. (11) also implies that
 the decoupling contour (14) continues into the coupled
phase at temperatures $T < T_m$, there  representing a crossover.
The first-order line 
 (14) must therefore {\it end} at
 the second-order melting line, $T = T_m$.
The melting curve then continues up in field along the latter
second-order line. 
Finally,  it is possible that a vestige of the second-order melting
transition at high fields $B_{\perp} > B_{\perp}^*$ persists
down into the low-field region in the form of a crossover
(see Fig. 1).

{\it Discussion.} The first-order flux-lattice melting line in clean
high-temperature superconductors is experimentally observed
to end at a multi-critical point.[2,3]  The phase diagram
proposed in Fig. 1 for the mixed phase of extremely type-II
layered superconductors also exhibits a multicritical point
at a temperature and field $T_0\sim T_m^{(2D)}$ and 
$B_0\sim B_{\perp}^*$. This point agrees qualitatively
with the high-$T_c$ phase diagram.[2,5]
Also, bulk pinning is observed to become relatively strong
in clean oxide superconductors  at low temperatures and high fields, 
$T < T_m^{(2D)}$ and $H_{\perp} > B_{\perp}^*$.[3]
The quasi-2D vortex lattice phase
that is identified
in Fig. 1 coincides  with this regime.
It   is a smectic (super) solid,[8]
and it  can thus adjust  better
to a random landscape of  point pins than a
rigid  vortex lattice.  Recent observations of muon spin resonance
in the mixed phase of clean  high-temperature superconductors
are consistent with  such a picture of
enhanced point pinning due to
 dimensional crossover.[23]  
We remind the reader that all effects
due to magnetic screening[24] have been 
completely neglected in the present theory.

In conclusion, the phase diagram proposed in Fig. 1 for the layered
$XY$ model with uniform frustration is strikingly similar to that
of the mixed phase in high-$T_c$ superconductors.[2,3]  It 
is perhaps more important to point out, however,
that the above duality analysis yields 	only {\it two}	
thermodynamic phases at weak coupling: a coupled superconductor (7) and a decoupled ``normal''
state (8).
This indicates that  neither
the Friedel scenario (decoupled superconducting
layers) nor the flux-line liquid state (coupled normal layers)
are thermodynamically possible in the absence of disorder.[1]

The author thanks M. Dodgson for correspondence and 
B. Rosenstein for discussions.

\vfill\eject
\centerline{\bf References}
\vskip 16 pt


\item {1.}  G. Blatter, M.V. Feigel'man, V.B. Geshkenbein, A.I. Larkin,
and V.M. Vinokur, Rev. Mod. Phys. {\rm 66}, 1125 (1994).

\item {2.} E. Zeldov, D. Majer, M. Konczykowski,
V.B. Geshkenbein, V.M. Vinokur, and H. Shtrikman,
Nature {\bf 375}, 373 (1995).

\item {3.} D.T. Fuchs, E. Zeldov, T. Tamegai, S. Ooi, M. Rappaport
and H. Shtrikman, Phys. Rev. Lett. {\bf 80}, 4971 (1998);
B. Khaykovich, E. Zeldov, D. Majer, T.W. Li, P.H. Kes, and
M. Konczykowski, Phys. Rev. Lett. {\bf 76}, 2555 (1996).


\item {4.}  T. Shibauchi, T. Nakano, M. Sato, T. Kisu, N. Kameda, N. Okuda,
S. Ooi, and T. Tamegai, Phys. Rev. Lett. {\bf 83}, 1010 (1999).

\item {5.} L.I. Glazman and A.E. Koshelev, Phys. Rev. B {\bf 43}, 2835 (1991).
 
\item {6.} L.L. Daemen, L.N. Bulaevskii, M.P. Maley and
J.Y. Coulter,  Phys. Rev. B {\bf 47}, 11291 (1993).
 
\item {7.} M. Feigel'man, V.B. Geshkenbein, and A.I. Larkin, 
Physica C {\bf 167}, 177 (1990).

\item {8.} E. Frey, D.R. Nelson, and D.S. Fisher,
Phys. Rev. B {\bf 49}, 9723 (1994).

\item {9.} B. Horovitz and T.R. Goldin, Phys. Rev. Lett. {\bf 80},
1734 (1998).

\item {10.} Y.H. Li and S. Teitel, Phys. Rev. B {\bf 47}, 359
(1993); {\it ibid} {\bf 49}, 4136 (1994).
 
\item {11.} A.E. Koshelev, Phys. Rev. Lett. {\bf 77}, 3901 (1996);
 Phys. Rev. B {\bf 56}, 11201 (1997).
 


\item {12.} J.P. Rodriguez, J. Phys. Cond. Matter {\bf 9}, 5117 (1997).
 
\item {13.} J.V. Jos\' e, L.P. Kadanoff, S. Kirkpatrick and
D.R. Nelson, Phys. Rev. B {\bf 16}, 1217 (1977).
 
\item {14.}  S.E. Korshunov, Europhys. Lett. {\bf 11}, 757 (1990).

\item {15.} J.P. Rodriguez, Europhys.  Lett. {\bf 31}, 479 (1995).

\item {16.} J.P. Rodriguez, ``Nature of Decoupling in the Mixed Phase
of Extremely Type-II Layered Superconductors'' (cond-mat/9906199).
 
\item {17.} J.P. Rodriguez, ``Theory for Decoupling in
the Mixed Phase of Extremely
Type-II Layered Superconductors'', (cond-mat/0006176).
  
\item {18.} S. Hikami and T. Tsuneto, Prog. Theor. Phys. {\bf 63}, 387 (1980);
see also S.R. Shenoy and B. Chattopadhyay, Phys. Rev. B {\bf 51}, 9129
(1995).

\item {19.} A. Polyakov, Phys. Lett. {\bf 72} B, 477 (1978).

\item {20.} D.R. Nelson and J.M. Kosterlitz, Phys. Rev. Lett. {\bf 39},
1201 (1977).

\item {21.} J.M. Kosterlitz and D.J. Thouless,
J. Phys. C {\bf 6}, 1181 (1973).

\item {22.} S.A. Hattel and J.M. Wheatley,
Phys. Rev. B {\bf 51}, 11951 (1995).

\item {23.} J.E. Sonier, J.H. Brewer, R.F. Kiefl, D.A. Bonn,
J. Chakhalian, S.R. Dunsiger, W.N. Hardy, R. Liang,
W.A. MacFarlane, R.I. Miller, D.R. Noakes, T.M. Riseman
and C.E. Stronach (cond-mat/9906449).

\item {24.} G. Blatter, V. Geshkenbein, A. Larkin and H. Nordborg,
Phys. Rev. B {\bf 54}, 72 (1996).





\vfill\eject
\centerline{\bf Figure Caption}
\vskip 20pt
\item {Fig. 1.}   
Shown is a  schematic phase diagram for the uniformly
frustrated $XY$ model (1) made up of a finite number of weakly-coupled
layers.  The cross-over transition that is represented by the
 horizontal dashed    line at low temperatures  
$T_{\rm J} < T < T_m^{(2D)}$ is broad (see ref. [17]).

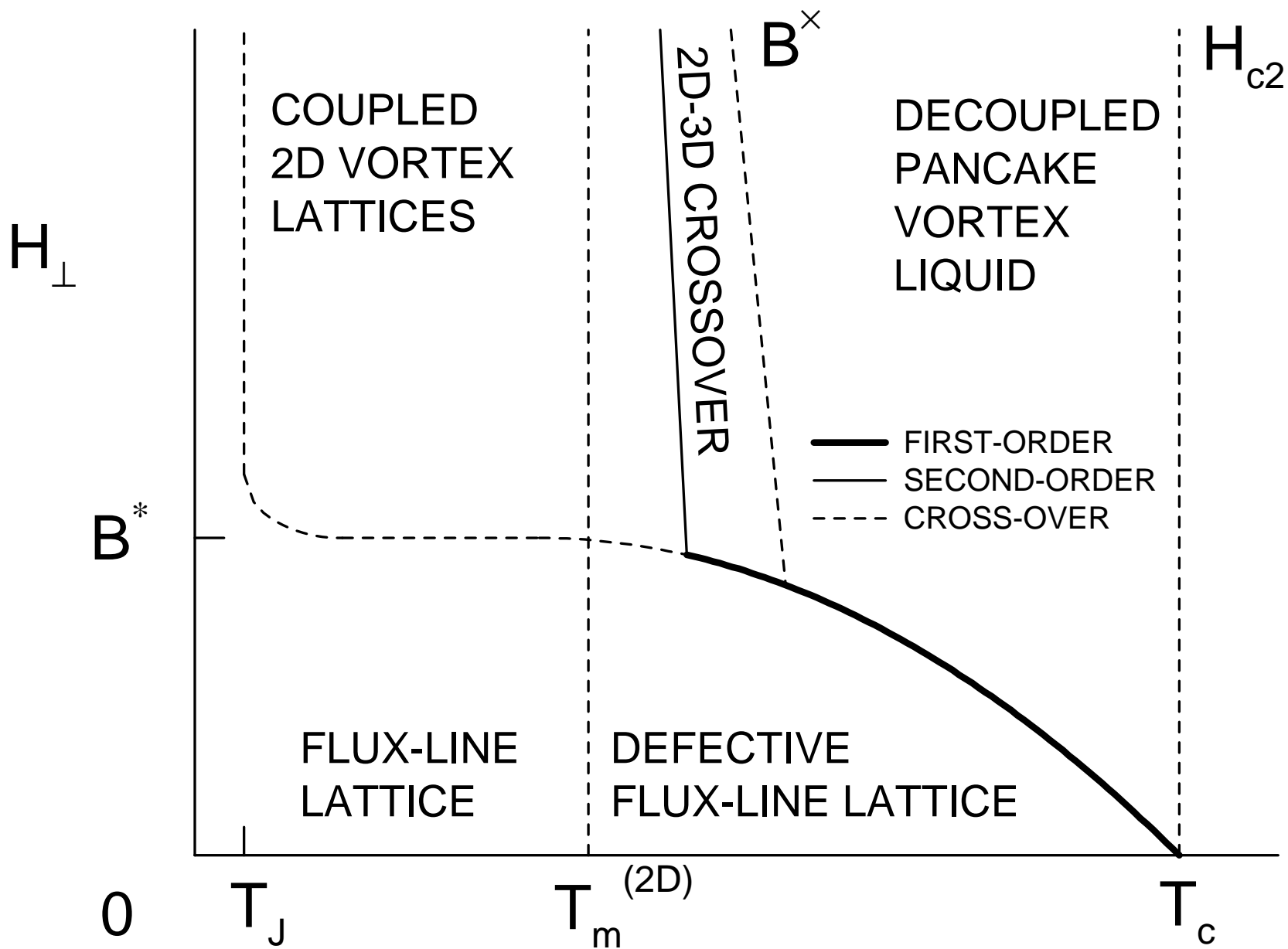
\end